# Macroscopic quantum correlation using coherence manipulations of polarization-path correlations of a continuous-wave laser

Byoung S. Ham[1,2]
[1]Center for Photon Information, School of Electrical Engineering and Computer Science, Gwangju Institute of Science and Technology, 123 Chumdangwagi-ro, Buk-gu, Gwangju 61005, South Korea
[2]Qu-Lidar, 123 Chumdangwagi-ro, Buk-gu, Gwangju 61005, South Korea
(Submitted on August 18, 2023; bham@gist.ac.kr)

**Abstract**
Quantum superposition is normally sustained in a microscopic regime governed by Heisenberg's uncertainty principle applicable to a single particle. Quantum correlation between paired particles implies the violation of local realism governed by classical physics. Over the last decades, quantum features have been implemented in various quantum technologies including quantum computing, communications, and sensing. Such quantum features are generally known to be impossible by any classical means. Here, a macroscopic quantum correlation is presented for coherence manipulations of polarization-path correlations of a continuous wave laser, satisfying the joint-parameter relation in an inseparable product-basis form. For the coherence control of the polarization-path correlation, a pair of electro-optic modulators is used in a noninterfering Mach-Zehnder interferometer for deterministic switching between paired polarization bases, resulting in the polarization product-basis superposition in a selective product-basis choice manner by a followed pair of acousto-optic modulators. This unprecedented macroscopic quantum feature opens the door to a new understanding of quantum mechanics beyond the microscopic regime for future classical optics-compatible quantum information.

**Introduction**
Since the dawn of quantum mechanics in the early 1900s, the complementarity theory between conjugate variables of a physical entity has become the holy grail in the present quantum science and technologies [1-5]. The uncertainty principle is regard to the minimum error in measurements of a single particle between its conjugate variables such as position and momentum, or energy and time. The wave-particle duality [2] is another expression of this uncertainty relation, where the phase term replaces the time. Thus, the conjugate variables can be used to compensate for the error product. In that sense, the wave-particle duality whose property is mutually exclusive is the ultimate case of the complementarity theory. This uncertainty relation is equivalent to the mathematics of Fourier transformation. In a Fourier series, a pulse is the direct result of many-wave interference in different wavelengths, representing the wave-particle duality, where their initial phases must coincide at a fixed difference. The coherence length of a commercial laser follows the wave-particle duality as the inverse of its linewidth (bandwidth) in the same many waves' ensemble decoherence. In quantum mechanics, this many-wave interference is basically equivalent to the particle nature via quantum superposition between them [6-9]. In the extreme case of a single photon for the particle nature with no energy error, no phase information can exist, and vice versa, satisfying the mutually exclusive wave-particle duality.

The quantum superposition of a single particle can be expressed with pure states of orthonormal bases [1,3,5]. The random probability between bases relates to the indistinguishable photon characteristics in quantum mechanics [5,10]. In a limited case with orthogonal polarization bases, however, the Heisenberg uncertainty principle has nothing to do with the determination of the photon characteristics. In general, the orthogonal basis relation of a single photon (particle) relates to the which-way information in an interferometer [10-15]. Wheeler's thought experiment of the delayed choice is for measurement-dependent photon characteristics of the wave-particle duality [10]. The quantum eraser based on Wheeler's delayed choice has demonstrated the violation of the cause-effect relation, i.e., the rule of thumb in classical physics [12-15]. Unlike quantum entanglement related to paired particles [16-22], the quantum eraser is basically for a single photon in an interferometric system via post-control of the quantum superposition. The nonlocal quantum feature belongs to the same physics of the violation of local realism in the quantum eraser. In that sense, quantum mechanics cannot be compatible with classical physics due to the violation of local realism.

Recently, challenges to the general understanding of quantum mechanics have been conducted using pure coherence optics to understand the century-long lasting quantum mystery of nonlocal correlation [23-25] as well as the violation of the cause-effect relation in the quantum eraser [26-28]. Regarding these new interpretations of quantum mechanics, a fundamental clue of the quantum feature has been found in the paired particles that Heisenberg's uncertainty principle does not have to be effective, as already agreed in the EPR discussions [16,17]. Instead, a strong foundation of quantum entanglement has been found in a definite phase relation between paired particles without violating quantum mechanics [23,24,29,30]. With the fixed phase relation between paired particles, observed quantum features have been well explained coherently for the Hong-Ou-Mandel effect [24,30,31], Franson correlation [24,32], and Bell inequality violation [25,33]. Regarding the quantum eraser [11-15], a coherence solution has also been derived from the experimental conditions of entangled photons via coherence manipulation of, e.g., polarization bases [25] and selective measurements [26]. Thus, the violation of the cause-effect relation is understood as a quantum illusion rooted in the intentional measurement-event loss. However, such coherence interpretations for the quantum features have been limited to a microscopic regime of a single particle. Here, a macroscopic quantum correlation is presented for the violation of local realism using coherence manipulations of a continuous wave (cw) laser. As a result, a classically excited macroscopic quantum entanglement is obtained from controlled quantum superposition between polarization-path correlations via selective measurements.

**Analysis**

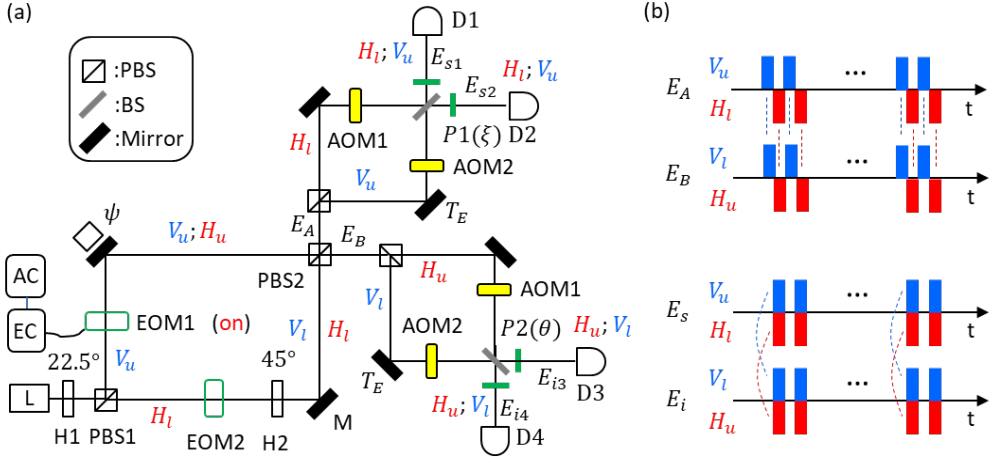

**Fig. 1. Macroscopic quantum correlation.** (a) Schematic of macroscopic quantum correlation. (b) Coherence manipulation-based pulse train. The red bars are by EOM switch-on. L: laser, AC: AOM controller, EOM: electro-optic modulator (EOM), EC: EOM controller, H1: 22.5°-rotated half-wave plate, H2: 45°-rotated half-wave plate, PBS: polarizing beam splitter, P: polarizer, D: photo-detector, PZT: piezo-electric transducer. $T_E$ is the pulse-to-pulse delay. AOM1 (AOM2) is for $+\Delta f$ ($-\Delta f$) frequency detuning from the center frequency $f_0$ of the cw laser.

Figure 1(a) shows a schematic of the macroscopic quantum correlation using a cw laser via coherence manipulations of the polarization-path correlation control in an MZI. To satisfy the wave nature, the linewidth of the cw laser is set to be very narrow, especially compared to the $\pm\Delta f$ of the acousto-optic modulators (AOMs) following the electro-optic modulators (EOMs). To provide random polarization bases, the cw light pulses pass through a 22.5°-rotated half-wave plate (HWP, H1). The polarizing beam splitter (PBS; PBS1) creates a polarization-path correlation inside the MZI: $|V\rangle_u|H\rangle_l$. By the following EOMs, the PBS1-determined polarizations are swapped in a 50% duty cycle in each MZI path, as shown in the two upper panels of Fig. 1(b). For the same-polarization correlation inside the MZI, a 45°-rotated HWP (H2) is inserted into the lower path: $|V\rangle_u|H\rangle_l \rightarrow |V\rangle_u|V\rangle_l$; $|H\rangle_u|V\rangle_l \rightarrow |H\rangle_u|H\rangle_l$. Thus, a series of alternative polarization-path correlations is provided, where $I_0$ is the light intensity and $\alpha$ is the phase between pairs: $|\Psi\rangle_M = I_0(|V\rangle_u|V\rangle_l + e^{i\alpha}|H\rangle_u|H\rangle_l)/2$; $k = u, l$. By PBS2, the same macroscopic relation is transferred to the MZI output paths.

As shown in the top two panels in Fig. 1(b), there is no temporal overlap between colored pairs. Thus, the macroscopic quantum superposition between these polarization-path correlations must be carefully considered. Unlike the single photon case [34] whose coherence length is much longer than the ensemble coherence [24], Fig. 1 has no superposition due to temporal extinction for orthogonal bases. To solve this critical matter, the output pulses are separated into polarization bases by the following PBS (PBS2) in each output port of the MZI, and recombined by a beam splitter via a $T_E$-delay adjustment, as shown in the two bottom panels of Fig. 1(b). Here, the time-delay adjustment is applied to the original pulse pairs in blue bars by $T_E$, where $T_E$ is a time period between EOM switching pulses. Now, the quantum superposition based on coherence between polarization-path correlations is satisfied. Moreover, the randomness in a quantum superposition between the product bases in the red and blue bars is also satisfied, where the entanglement relation $|\Psi\rangle_{si} = I_0(|V\rangle_u|V\rangle_l + |H\rangle_u|H\rangle_l)/2$ is targeted to achieve for measurements. The polarization-path correlations in $|\Psi\rangle_{si}$ are mutually exclusive in the time domain for a microscopic regime [25].

In Fig. 1, the polarization basis product between two detectors, e.g., D1 and D3 (D2 and D4), a total of four product bases result in the 2x2 tensor products. For $|\Psi\rangle_{si}$, thus, a total 50% of product bases needs to drop. For this, a gated heterodyne detection method is adopted via AOM modulated signals. The AOM1 in each party is for $+\Delta f$ addition to $H_u$ and $V_u$, while AOM2 is for $-\Delta f$ to $H_l$ and $V_l$. To keep coherence among them, the EOMs and AOMs must be synchronized. Due to the synchronously paired polarization bases, thus, the exclusive nature between colored polarization-product bases is now satisfied in a macroscopic regime. Finally, by the action of polarizers $(\xi, \theta)$, the quantum eraser is conducted even for a cw regime [26,35].

Due to the macroscopic quantum feature introduced in Fig. 1, the kets are replaced by electric field vectors, as usual in coherence optics:

$$\boldsymbol{E}_A = \frac{E_0}{2}\left(-\hat{V}_u e^{i\psi} + \hat{H}_l\right), \tag{1}$$

$$\boldsymbol{E}_B = \frac{iE_0}{2}\left(\hat{V}_l + \hat{H}_u e^{i\psi}\right), \tag{2}$$

where $E_0$ is the amplitude of an input pulse to the MZI. In Eqs. (1) and (2), the hat represents a unit vector of the electric fields. As a result, the corresponding mean intensities become $\langle I_A \rangle = \langle I_B \rangle = I_0/2$ due to the orthogonal polarization bases [36,37], resulting in the same particle nature of distinguishable photon characteristics in quantum mechanics [14,15]. In Eqs. (1) and (2), the time separation between orthogonally polarized pulses is the direct reason of no fringes.

To solve this incoherence matter between the EOM switched and original pulse sets in Eqs. (1) and (2), the original pulse sets need to be delayed in the following MZIs. In addition, AOMs are inserted for the heterodyne detection, too. By the polarizer inserted in each output port of the second MZIs in Fig .1(a), Eqs. (1) and (2) turn out to be coherent due to the allowance of the cross-polarization products by the polarization projection on the polarizer:

$$\boldsymbol{E}_{s1} = \frac{iE_0}{2}\left(-\hat{V}_u e^{i\varphi} sin\xi + \hat{H}_l cos\xi\right), \tag{3}$$

$$\boldsymbol{E}_{s2} = \frac{E_0}{2}\left(\hat{V}_u e^{i\varphi} sin\xi + \hat{H}_l cos\xi\right), \tag{4}$$

$$\boldsymbol{E}_{i3} = \frac{-E_0}{2}\left(\hat{V}_l sin\theta + \hat{H}_u e^{i\varphi} cos\theta\right), \tag{5}$$

$$\boldsymbol{E}_{i4} = \frac{iE_0}{2}\left(-\hat{V}_l sin\theta + \hat{H}_u e^{i\varphi} cos\theta\right), \tag{6}$$

where $\xi$ and $\theta$ are the rotation angles of the inserted polarizers from the horizontal axis to the counterclockwise direction [26]. Here, the unit vectors are just to indicate the origins of the light pulses. The time delay of the vertically polarized light pulses is included in $\varphi$, where $\varphi = \psi + \zeta(T_E) + 2\Delta f\tau$, where $\zeta$ and $\psi$ are much more sensitive compared to $\Delta f\tau$, but actively controllable. Thus, the quantum eraser is $\Delta f\tau$ dominated.

Compared with uniform intensities of $\langle I_A \rangle$ and $\langle I_B \rangle$, the locally measured mean intensities of Eqs.(3)-(6) result in interference fringes [12-15,26,34,35]:

$$\langle I_{s1}\rangle = \langle I_{i4}\rangle = \frac{I_0}{4}\langle 1 - sin2\xi cos\varphi\rangle, \tag{7}$$

$$\langle I_{s2}\rangle = \langle I_{i3}\rangle = \frac{I_0}{4}\langle 1 + sin2\theta cos\varphi\rangle. \tag{8}$$

For the ensemble decoherence of photon pairs from such as spontaneous parametric down-conversion (SPDC) [31-34], the $\cos\varphi$ term in Eqs. (7) and (8) is much less sensitive to the MZI path-length difference of the usual first-order intensity correlation. In that sense, the AOMs could be scanned for a certain bandwidth comparable to $\Delta f$, but not necessary. Under active stabilization at near $\tau \sim 0$, Eqs. (7) and (8) become the function of polarizers only, satisfying the quantum eraser.

Like coincidence measurements in a single photon regime [19-22,32-34], the mean output intensity products in Fig. 1(a) can also be represented in the same way due to the randomness of the polarization-path correlations. For this, selective measurements are conducted by a gated heterodyne detection [25]:

$$\langle R_{s1i3} \rangle = \langle R_{s2i4} \rangle = \frac{I_0^2}{16} \langle \left(-\hat{V}_u e^{i\varphi} \sin\xi + \hat{H}_l \cos\xi\right)\left(\hat{V}_l \sin\theta + \hat{H}_u e^{i\varphi} \cos\theta\right)(cc) \rangle,$$
$$= \frac{I_0^2}{16} \langle \left(-\hat{V}_u \hat{V}_l \sin\theta \sin\xi + \hat{H}_l \hat{H}_u \cos\theta \cos\xi\right)(cc) \rangle,$$
$$= \frac{I_0^2}{16} \langle \cos^2(\xi + \theta) \rangle. \quad (9)$$

Interestingly, Eq. (9) is $\varphi$-independent, as observed in nonlocal quantum correlations in a microscopic regime [19-22,32-34]. Here, it should be noted that only the same polarization-product bases $\hat{V}_u \hat{V}_l$ and $\hat{H}_l \hat{H}_u$ are allowed in Eq. (9) by the action of the gated heterodyne detection. Although the exclusive nature of the polarization-path correlations is provided by EOMs, the quantum eraser by $\xi$ or $\theta$ via $T_E$ delay adjustment retrieves the wave nature, resulting in the allowance of $\hat{V}_u \hat{H}_u$ and $\hat{V}_l \hat{H}_l$, too. Removing these product bases from the measurements is the purpose of the AOM-based heterodyne detection [25]. Unlike the conventional understanding of quantum mechanics, thus, Eq. (9) for Fig. 1 results in the macroscopic quantum correlation in a joint-phase relation of local parameters $\xi$ and $\theta$. The origin of the macroscopic quantum feature in Eq. (9) is in the coherently manipulated quantum superposition between the polarization-path correlations by EOMs and the AOM-based selective measurement via delay adjustment. The violation of local realism in Eq. (9) is determined by the path length between PBS1 and the polarizers which should be beyond the light cone by definition (see Discussion) [19].

**Discussion**

*Quantum illusion vs selective measurements*
Like coincidently provided entangled photon pairs [24,32], the same feature of reduced polarization-product bases are coherently provided in Fig. 1 for $\hat{V}_u \hat{H}_l$ and $\hat{V}_l \hat{H}_u$, as shown in Eq. (9). The goal of synchronized EOMs and AOMs is to provide such selective polarization-product bases for quantum superposition between them. In the Bell inequality violations, similar reduced measurements are conducted by polarization projection using the quantum eraser [33,34]. The selective measurements are for 50% measurement event loss at least. Thus, the quantum correlation can be viewed as a quantum illusion like a rainbow color filtered out from the sun lights through a prism or diffraction.

*Nonlocal realism*
The conventional interpretation of nonlocal quantum features is for statistical ensemble without any phase relation between measured events. In Fig. 1, such independence is provided by EOMs and $T_E$-based temporal delay. On the other hand, quantum superposition between product bases is provided by quantum coherence between paired particles in each measurement event. This coherence between photons in each measurement event is inherent and cannot deteriorate in space within the given coherence. This critical difference in conceptual understandings originates in the mutually exclusive wave-particle duality, where a photon's particle (wave) nature must neglect the phase (energy) information. For the particle nature, thus, the local realism between independent particles must be confined by the spatial range determined by a force acting on them. On the contrary, the local realism of the wave nature-based interpretation is determined by the coherence length as a predetermined nature. For the ultimate case with monochromatic light, the coherence length is infinite, even in the case of the EOM switching in an adiabatic way. In the present analysis, no force is related to the coherence (wave nature) because there is nothing to change the given basis status. Thus, the measurements of the polarization-path correlated paired pulses are just to check the given states at a given measurement condition of, e.g., space-like separations. As a result, nonlocal realism might have no practical meaning to the present coherence interpretation.

*Ensemble effect*

The HOM effects do not show the φ-dependent interference fringes as in the coherence optics. This is the direct result of many-wave interference [31]. Instead, HOM effects show ensemble-based beating fringes [30]. In the Bell inequality violation based on the quantum eraser, no such many-wave interference effect exists, as derived in Eq. (9) for the φ-independence. Such a no-ensemble effect has also been successfully analyzed for the Franson correlation [24]. The difference between the HOM effect and Bell inequality violation is in the selective measurements by either polarization projection [30], coincidence measurements [24], or gated heterodyne detection [25]. Regarding Fig. 1(b), AOMs can be either fixed at $\pm \Delta f$ or scanned for a given bandwidth [25].

**Conclusion**

A macroscopically excited nonlocal quantum feature was proposed and analyzed for a cw laser using coherence manipulations of polarization-path correlations by EOMs and a gated heterodyne detection by AOMs. Unlike common methods of entangled pairs such as SPDC nonlinear optics limited to a microscopic regime, the proposed nonlocal quantum feature was for a macroscopic regime that has never been expected or discussed before, which is contradictory to the common understanding. For the coherence manipulations of quantum superposition between random polarization-path correlations, a coherence time delay was used to temporally separate product pairs. By the action of EOMs, polarization-path correlations were alternatively swapped, resulting in the macroscopic polarization-product bases for the intensity products between space-like separated The function of AOMs was to add frequency correlation to the polarization-path correlation for the preparation of 50% reduced selective measurements by a gated heterodyne detection. Unlike our common understanding of quantum mechanics, thus, the macroscopic quantum feature was coherently excited using a cw laser light. This contradictory result may open the door to a new world of quantum information science for potential applications of unprecedented quantum technologies compatible with classical physics.

**Funding:** This research was supported by the MSIT (Ministry of Science and ICT), Korea, under the ITRC (Information Technology Research Center) support program (IITP 2023-2021-0-01810) supervised by the IITP (Institute for Information & Communications Technology Planning & Evaluation). BSH also acknowledges that this work was also supported by GIST GRI-2023.